\documentclass[9pt]{article}  \usepackage{times}
\usepackage{graphicx, textgreek}

\usepackage{color}

\usepackage[round,numbers,sort&compress]{natbib} 

\usepackage{amsmath}
\usepackage{nccmath}   
\usepackage{amssymb}
\usepackage{stackrel}
\usepackage{chemarrow}
\usepackage{graphicx}
\usepackage{textgreek}

\usepackage{booktabs}
\usepackage{dcolumn}
\usepackage{rotating}
\usepackage{multirow}

\renewcommand\appendix{\par
	\setcounter{section}{0}
	\setcounter{subsection}{0}
	\setcounter{figure}{0}
	\setcounter{table}{0}
	\renewcommand\thesection{Appendix \Alph{section}}
	\renewcommand\thefigure{\Alph{section}\arabic{figure}}
	\renewcommand\thetable{\Alph{section}\arabic{table}}
}

\begin{document}

	\twocolumn[{\LARGE \textbf{Damped physical oscillators, temperature and chemical clocks\\*[0.2cm]}}
	{\large Thomas Heimburg$^\ast$\\*[0.1cm]
		{\small Niels Bohr Institute, University of Copenhagen, Blegdamsvej 17, 2100 Copenhagen \O, Denmark}\\*[-0.1cm]
		
		{\normalsize \textbf{ABSTRACT}\hspace{0.5cm} The metaphor of a clock in physics describes near-equilibrium reversible phenomena such as an oscillating spring. It is surprising that for chemical and biological clocks the focus has been exclusively on the far-from-equilibrium dissipative processes. We show here that one can represent chemical oscillations (the Lotka-Volterra system and the Brusselator) by equations analogous to Onsager's phenomenological equations when the condition of the reciprocal relations, i.e. the symmetry in the coupling of thermodynamic forces to fluxes is relaxed and antisymmetric contributions are permitted. We compare these oscillations to damped oscillators in physics (e.g., springs, coupled springs and electrical circuits) which are represented by similar equations. Onsager's equations and harmonic Hamiltonian systems are shown to be limiting cases of a more general formalism. 
		
		The central element of un-damped physical oscillations is the conservation of entropy which unavoidably results in reversible temperature oscillations. Such temperature oscillations exist in springs and electrical LC-circuits, but have among others also been found in the oscillating Belousov-Zhabotinsky reaction, in oscillations of yeast cells, and during the nervous impulse. This suggests that such oscillations contain reversible entropy-conserving elements, and that physical and chemical clocks may be more similar than expected.
			\\*[0.3cm] }}
	\noindent\footnotesize{\textbf{Keywords:} Clocks, chemical oscillations, temperature oscillations, linear nonequilibrium thermodynamics\\*[0.1cm]}
	\noindent\footnotesize {$^{\ast}$corresponding author, theimbu@nbi.ku.dk. }\\
	\vspace{0.3cm}
	]

	\normalsize

\section{Introduction}
\label{introduction}

Oscillating chemical reactions are fascinating and important phenomena. Their emergence is so surprising that at first they were not believed to be possible. This is easily understood since chemical reactions are considered as over-damped systems. Thus, it is difficult to imagine the origin of such oscillations. Besides famous examples such as the Belousov-Zhabotinsky reaction \cite{Zaikin1970} and the Briggs-Rauscher reaction \cite{Briggs1973} with oscillating bromate and iodate concentrations, oscillating reactions are also believed to be important in biology \cite{Rensing2001}, e.g., in the glycolytic cycle \cite{Dutt1993}, reproduction cycles of cells \cite{Teusink1996, Thoke2015, Thoke2018}, the heart beat \cite{Krinsky1978, Lakatta2010} and in circadian rhythms \cite{Vitaterna2001, Albrecht2012}.

There exist non-physical examples of coupled equation systems that exhibit such oscillations \cite{Lotka1925, Kerner1957}. The predator-prey systems, e.g., the populations of lynx and snowshoe hare given the abundance of food for the hare, lead to oscillations of the respective species that are described by a set of coupled rate equations called the Lotka-Volterra equations.\footnote{It has been questioned whether the Lotka-Volterra scheme describes the hare\slash lynx dynamics correctly because the amount of food may not stay constant \cite{Stenseth1997}.} Since there exists a phase shift between a decline in the hare population and the population of lynx (who can survive for some time without food), the system contains some ``inertia''. In analogy to the predator-prey scheme, one has proposed sets of nonlinear coupled rate equations involving the autocatalytic generation of oscillating reaction intermediates \cite{Nicolis1971, Nicolis1971b, Kondepudi1998}. Such systems of equations typically contain only concentrations of reaction partners and do not consider any other thermodynamic variables.

A seemingly different system is the oscillation of a spring. For undamped oscillations, kinetic energy is reversibly transformed into potential energy and vice versa. It is clear that a predator-prey scheme is qualitatively different because hares do not eat lynxes, and there is no reversible conversion between the two species. For this reason we wish to distinguish these two kinds of oscillations. When the Lotka-Volterra sche\-me is applied to chemical reactions, the reaction has the form $S \rightarrow P$ meaning that a substrate $S$ is converted into a product $P$ in an irreversible manner. The way that the reactions are written suggests a resemblance with the predator-prey scheme and is distinct from Hamiltonian systems. The mathematical treatment consists of three coupled differential equations in which two intermediates oscillate. This reaction scheme in fact leads to oscillations of intermediates that resemble those found in experiments. Another example is the Belousov-Zha\-botinsky-reaction. A reaction scheme for describing this reaction (called the Field-K\"or\"os-Noyes (FKN) reaction) has been proposed by \cite{Field1972} and \cite{Field1974}. A well-known minimalistic reaction scheme of similar nature is the Brusselator consisting of four coupled rate equations \cite{Nicolis1971, Kondepudi1998}. The overall reaction in this scheme is $S_1 + S_2 \rightarrow P_1 + P_2$, i.e., two substrates are converted into two products. As with Lotka-Volterra equations, there exist two reaction intermediates that oscillate.

The physical origin of such oscillations is not immediately clear. Oscillations in physics most often involve the interplay between inertial forces and those arising from potential gradients. In the absence of inertia, the origin of chemical oscillations seems more obscure. Nevertheless, it has been found that the Lotka-Volterra equations can be rewritten to resemble Hamilton's equations of motion \cite{Kerner1957, Kerner1959, Kerner1964, Nutku1990, Plank1995}. As suggested above, this similarity appears to be of a mathematical rather than a physical nature. In this paper we wish to explore the possibility that this resemblance is not fortuitous but that chemical oscillations in fact represent physical oscillations with inertial elements.

Non-equilibrium phenomena are usually treated using Onsager's phenomenological equations \cite{Onsager1931b}. One of the assumptions leading to these equations is that the entropy can be expanded as a harmonic function around the equilibrium state. It describes the probability of fluctuations at constant temperature. Thermodynamic forces $X_i$ are the gradients of this function, and thermodynamic fluxes $J_i$ (i.e., the flux of an extensive variable from a non-equilibrium state towards equilibrium) are assumed to be linear functions of forces. Simple examples are Ohm's law (current is proportional to voltage) and Stoke's law (velocity of a sphere in a viscous medium is constant for a constant applied force). The phenomenological equations can be written as $\underline{J}=\underline{\underline{L}}^S\;\underline{X}$, where $\underline{J}$ is a vector containing the thermodynamic fluxes, $\underline{X}$ is a vector contain thermodynamic forces, and $\underline{\underline{L}}^S$ is a symmetric matrix. The symmetry in this matrix is known a Onsager's reciprocal relations. It has been justified by the independence of the thermal fluctuations on the direction of time and assumes the complete absence of inertial forces, i.e., they describe over-damped systems. Onsager's equations do not contain a mechanism that allows for oscillations and therefore it is assumed that chemical oscillations such as the Belousov-Zhabotinsky reaction must be of far-from-equilibrium nature due to higher-order effects \cite{Nicolis1971}.

Since the equilibrium of a spring is the resting position, oscillating springs are also non-equilibrium systems. We show here that one can generalize Onsager's phenomenological e\-quations in a simple manner such that they contain inertia if one drops the constraint of the reciprocal relations. This formalism allows to describe damped oscillating processes with an equation scheme similar to that of Onsager. One can find simple equations for damped springs, chains of springs, electrical oscillators and chemical oscillations that are written in a thermodynamic language but contain inertia. If undamped, these oscillations conserve entropy, and as a direct consequence one finds oscillations in temperature. Temperature changes are familiar from the adiabatic compression of materials (e.g., gases but also real springs \cite{Heimburg2021}) but also for the adiabatic charging of capacitors \cite{Mischenko2006, Scott2011, Crossley2016, Janssen2017}. As we have shown in \cite{Heimburg2017}, there exist analogues of the inertial mass in systems that are not of mechanical nature, e.g., the inductance of coils in an electrical LC-circuit that lead to oscillatory behavior.

We will argue that the main indicator that distinguishes the mathematical oscillations of coupled rate equations such as Lotka-Volterra or the Brusselator, from those of Hamiltonian systems containing inertial forces is the oscillation in temperature. While such oscillations are unexpected in the former systems, they are unavoidable in the latter systems. In fact, temperature oscillations have been found in the Belousov-Zhabotinsky reaction \cite{Franck1971, Franck1978, Boeckmann1996} but also in metabolic oscillations of yeast cells \cite{Teusink1996, Thoke2018} and temperature pulses during the action potential in nerve axons \cite{Abbott1958, Howarth1968, Howarth1975, Ritchie1985, Heimburg2021}. We will explore this in the cases of the damped oscillations in coupled springs, electrical circuits containing capacitors, inductors and resistors as well as for the example of the Lotka-Volterra and Brusselator oscillators. Finally, we will argue that Onsager's phenomenological equations and Hamilton's equations of motion for a harmonic potential are two limiting cases of a more generic set of equations that includes both inertia and friction and can naturally lead to oscillations in any thermodynamic system.


\section{Phenomenological equations}
\label{phenomenologicalequations}

Non-equilibrium thermodynamics is usually applied to closed systems in which the total energy is conserved and only the entropy increases. The entropy is approximated as a harmonic function
\begin{equation}
	S=S_0-\frac{1}{2}\sum_{ij} g_{ij}\alpha_i\alpha_j
\label{eq:ent3.1}
\end{equation}
where $\alpha_i=\xi_i-\xi_{i,0}$ is the variation of the extensive variable $\xi_i$ around its equilibrium position $\xi_{i,0}$, and $g_{jk}$ is a symmetric real positive definite matrix representing the metric of the entropy potential. Onsager's phenomenological equations describe the coupling between thermodynamic fluxes and forces. They are given by

\begin{equation}
J_i=L_{i1}^SX_1+L_{i2}^SX_2+...=\sum_i L_{ij}^SX_j \qquad i,j=1,...,n\\
\label{eq:ent3.2}
\end{equation}
where

\begin{align}
J_i&=\frac{\partial \alpha_i}{\partial t}&\mbox{thermodynamic flux}\nonumber\\
X_j&=\frac{\partial S}{\partial \alpha_j}=-\sum_k g_{jk}\alpha_k&\mbox{thermodynamic force}\nonumber\\
L_{ij}^S&=L_{ji}^S&\mbox{reciprocal relations} \;.\nonumber
\end{align}

In the case of two fluxes and two forces, Onsager's equations are given by

\begin{equation}\label{eq:ent3.6a}  
\left(\begin{array}{c}             
	\dot{\alpha}_1	\\
	\dot{\alpha}_2	
	\end{array}\right)=
	-
\left(\begin{array}{cc}  
L_{11}^S & L_{12}^S \\
	L_{21}^S & L_{22}^S 	
	\end{array}\right)
\left(\begin{array}{cc}       
	g_{11} & g_{12} \\
g_{21} & g_{22} 	
	\end{array}\right)
\left(\begin{array}{c}             
	\alpha_1	\\
	\alpha_2	
	\end{array}\right) \;,
\end{equation}

\noindent where both $\underline{\underline{L}}^S$ and $\underline{\underline{g}}$ are symmetric matrices. Simple physical laws of this nature are Ohm's law where the flux in charges is proportional to an electrical potential difference, and Stokes' law which describes the constant velocity of a particle pulled with constant force through a viscous fluid. In more complex situations, the fluxes and forces of non-conjugated variables can couple and, for instance, describe the thermoelectric effect or thermal osmosis. These systems are characterized by over-damping, and inertial forces are neglected.

It has been shown that the entropy production $\sigma=\partial s/\partial t$, i.e., the change of entropy density in a closed system with time is given by the product of fluxes and conjugated forces \cite{Kondepudi1998}

\begin{equation}
\sigma=\sum_i J_i\cdot X_i=\sum_{ij}L_{ij}^SX_i X_j \;.
\label{eq:ent3.3}
\end{equation}
In \cite{Heimburg2017} we discussed the role of antisymmetric contributions to the matrix $\underline{\underline{L}}$. Any $n \times n$ matrix can be written as the sum of a symmetric and an antisymmetric matrix \cite{Heimburg2017}, $L_{ij}=L_{ij}^S+L_{ij}^A$, where $L_{ij}^S=(L_{ij}+L_{ji})/2$ and $L_{ij}^A=(L_{ij}-L_{ji})/2$. The entropy production could generally be written as

\begin{equation}
\sigma=\sum_{ij}\left(L_{ij}^S + L_{ij}^A\right) X_i X_j \;,
\label{eq:ent3.4}
\end{equation}
which is identical to eq. (\ref{eq:ent3.3}) because all contributions from the antisymmetric matrix will cancel.

We have shown in \cite[]{Heimburg2017} that harmonic oscillators can be written in the form

\begin{equation}
\underline{J}=\underline{\underline{L}}^A \underline{X}=-\underline{\underline{L}}^A \underline{\underline{g}}\underline{\alpha} \;.
\label{eq:ent3.5}
\end{equation}

\begin{figure}[htbp]
\centering
\includegraphics[width=225pt,height=90pt]{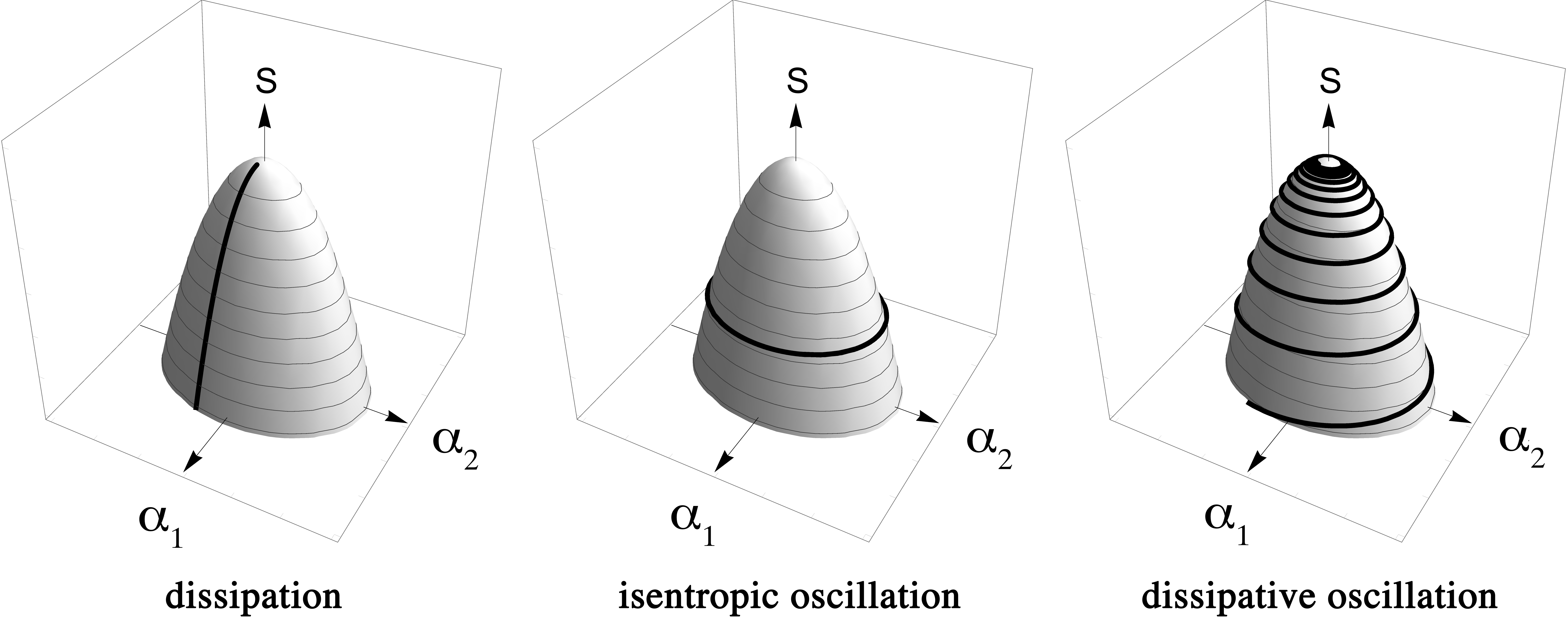}
\caption{Entropy changes in a system of two variables, $\alpha_1$ and $\alpha_2$. Left: Entropy increase in the limit of high friction (no inertia). Center: Entropy change in the absence of friction dominated by inertia. Right: Entropy changes in the presence of both friction and inertia. Adapted from \cite{Heimburg2017}.}
\label{figure_01c}
\end{figure}

With both matrices ($\underline{\underline{L}}^S$ and $\underline{\underline{L}}^A$) present, one obtains a damped oscillation.

\begin{equation}
\underline{J}=\left(\underline{\underline{L}}^S+\underline{\underline{L}}^A\right)\underline{X}=-\left(\underline{\underline{L}}^S+\underline{\underline{L}}^A\right)\underline{\underline{g}}\underline{\alpha} \;.
\label{eq:ent3.5b}
\end{equation}
This is shown schematically in Fig. \ref{figure_01c} (right).

One purpose of the present paper is to show that oscillating Hamiltonian systems with inertia in the absence and presence of friction can be written in a form that is analogous to Onsager's equations, but contain oscillations in temperature if springs or capacitors are considered as thermodynamic systems.

Consider the case of an undamped oscillation as shown in Fig \ref{figure_01c} (center) with two variables $\alpha_1$ and $\alpha_2$. We can write \cite{Heimburg2017}

\begin{equation}\label{eq:ent3.6}  
\left(\begin{array}{c}             
	\dot{\alpha}_1	\\
	\dot{\alpha}_2	
	\end{array}\right)=
	-
\left(\begin{array}{cc}  
	0 & L_{12}^A \\
	-L_{12}^A & 0 	
	\end{array}\right)
\left(\begin{array}{cc}       
	g_{11} & g_{12} \\
	g_{21} & g_{22} 	
	\end{array}\right)
\left(\begin{array}{c}             
	\alpha_1	\\
	\alpha_2	
	\end{array}\right) \;.
\end{equation}

The antisymmetric matrix enforces a coupling between the two variables $\alpha_1$ and $\alpha_2$:

\begin{equation}
	\alpha_2=-\frac{1}{L_{12}^A g_{22}}\frac{d\alpha_1}{dt} \;.
\label{eq:ent3.7}
\end{equation}

Due to eq. (\ref{eq:ent3.1}), The entropy is given by

\begin{equation}
	S-S_0=-\frac{1}{2}g_{11}\alpha_1^2-g_{12}\alpha_2\alpha_2-\frac{1}{2}g_{22}\alpha_2^2 \;,
\label{eq:ent3.8}
\end{equation}
which together with eq. (\ref{eq:ent3.6}) yields \cite{Heimburg2017}

\begin{align}
\frac{\partial S}{\partial \alpha_1}&=-\frac{1}{L_{12}^A}\dot{\alpha}_2\nonumber\\
\frac{\partial S}{\partial \alpha_2}&=+\frac{1}{L_{12}^A}\dot{\alpha}_1 \;.
\label{eq:ent3.9}
\end{align}
This is analogous to Hamilton's equations of motion. The oscillating system possesses an eigenfrequency $\omega$ with $\omega^2=(L_{12}^A)^2 det(\underline{\underline{g}})$.


\section{Entropy and temperature of springs}
\label{entropyandtemperatureofsprings}

The first law of thermodynamics for a spring is given by

\begin{equation}
dE=TdS  - Fdx  \;,
	\label{eq:ent1.1}
\end{equation}

\noindent where $E$, $S$ and $x$ are the extensive quantities internal energy, entropy, and position. $T$, and $F$ are the intensive quantities temperature and force, and $x$ is the relative displacement of the spring position from its equilibrium value. At constant entropy, $dE=-Fdx$. The force is given by $F=-Kx$, where K is the adiabatic compression modulus of the spring, $K=-(\partial F/\partial x)_S$, also called the \emph{`spring constant'}.

\begin{figure}[htbp]
\centering
\includegraphics[width=169pt,height=91pt]{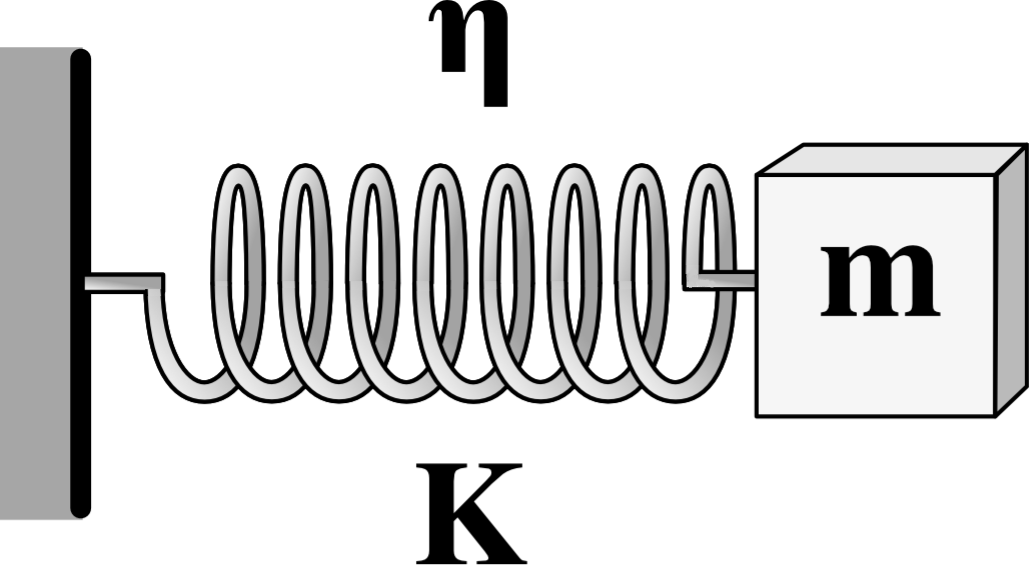}
\caption{Spring with spring constant $K$ connected to a mass $m$ with a friction coefficient $\eta$.}
\label{figure_01}
\end{figure}

Constant entropy implies that
\begin{equation}
\underbrace{T\left(\frac{\partial S}{\partial T}\right)_x}_{c_x} dT+T\left(\frac{\partial S}{\partial x}\right)_T dx=0 \;.
	\label{eq:ent1.2}
\end{equation}
Here, $c_x=T(\partial S/\partial T)_x$ is the heat capacity of the spring at constant extension, and $T$ is the temperature of the spring.

In \cite{Heimburg2021} we showed that, due to a Maxwell relation,

\begin{equation}
(\partial S/\partial x)_T=(\partial F/\partial T)_x\equiv g \;.
	\label{eq:ent1.2b}
\end{equation}
This implies that we can measure the isothermal dependence of the entropy on position by determination of the temperature-dependence of the force in a spring at constant position \cite{Heimburg2021}.

We define $x=0$ as the equilibrium position of the spring where its entropy is at maximum. Using eqs. (\ref{eq:ent3.1}) and (\ref{eq:ent1.2b}) we find that
\begin{equation}
	S=S_0-\frac{1}{2}g x^2 + ...\qquad \mbox{for}\; T=\mbox{const}\;.
	\label{eq:ent1.3}
\end{equation}
where $g$ is an isothermal modulus, and $(\partial S/\partial x)_T=-g\cdot x$ is a thermodynamic force. For adiabatic compression we find

\begin{equation}
	dS=\frac{c_x}{T} dT-gxdx=0
	\label{eq:ent1.4}
\end{equation}
and for small $x$

\begin{equation}
	\Delta T=\frac{T_0}{2c_x}gx^2 \;,
	\label{eq:ent1.5}
\end{equation}
where $\Delta T$ is the change of the temperature between a relaxed spring at $T_0$ and the compressed spring at extension $x$. The temperature is at minimum in the relaxed state of the spring, and at maximum in the most compressed or expanded state.

Let us compress a spring in the absence of friction. Instead of using an external force to compress that spring, we apply an inertial force. Both energy and entropy are conserved:

\begin{equation}
E=\frac{1}{2}Kx^2+\frac{p^2}{2m} =\mbox{const.}
\label{eq:ent2.1}
\end{equation}

\begin{equation}
S=-\frac{1}{2}gx^2+\frac{c_x\Delta T}{T_0}=\mbox{const.}
\label{eq:ent2.2}
\end{equation}

\begin{figure}[htbp]
\centering
\includegraphics[width=225pt,height=225pt]{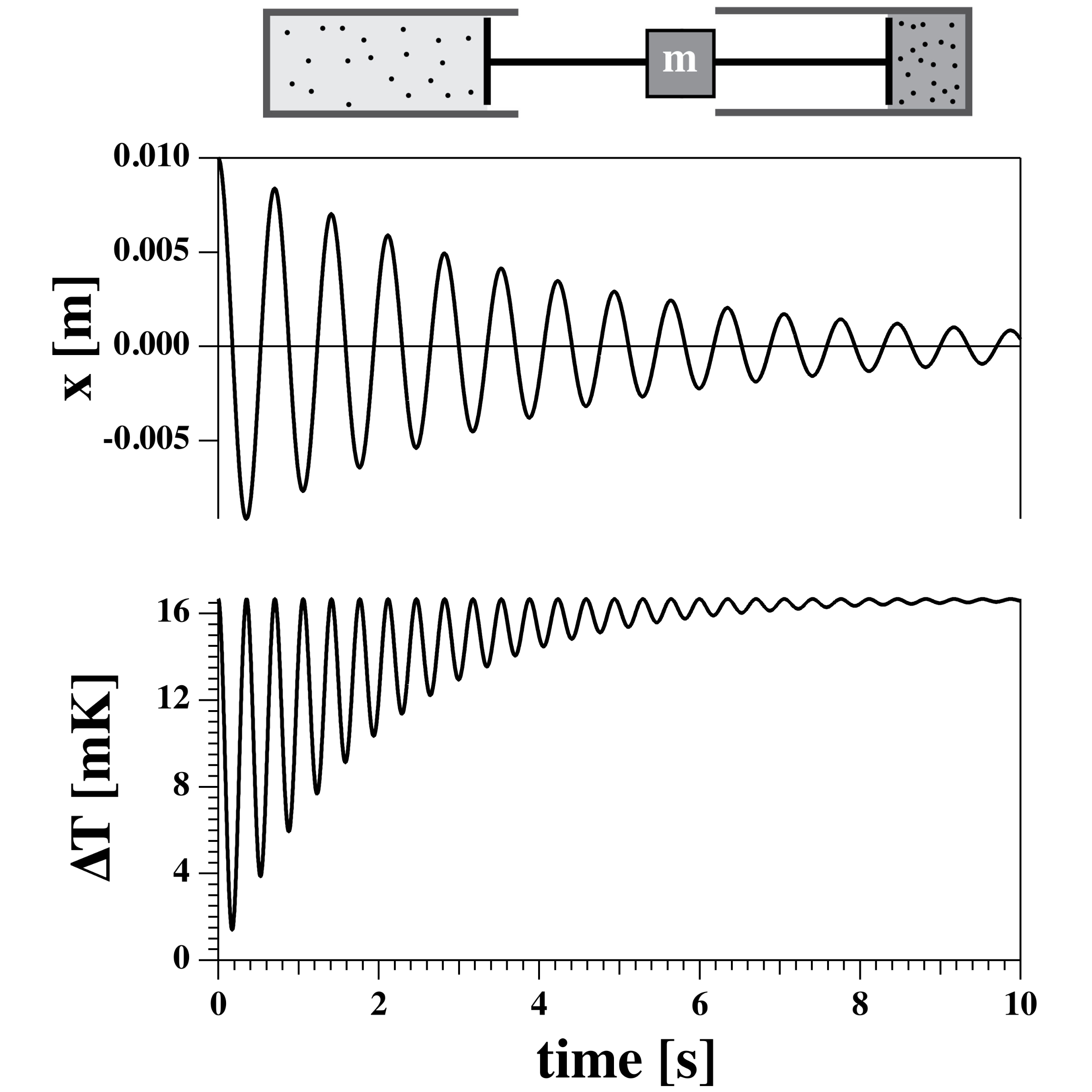}
\caption{Positional oscillations of an inertial mass $m$ coupled to two pistons filled with an ideal gas in the presence of damping. Each container contains 1 mol of particles, $T_0=300$ K, $x_0=1$ m (length of each container in equilibrium), $m=100$ kg, and the friction coefficient is $\eta=50$ Ns\slash m. The dissipation of energy leads to an exponential increase of the mean temperature of the two gas pistons, with superimposed decaying oscillations.}
\label{figure_00}
\end{figure}

Thus we find
\begin{equation}
\frac{c_x\Delta T}{T_0}+\frac{gp^2}{2mK}=\mbox{const.}
\label{eq:ent2.3}
\end{equation}

and

\begin{equation}
\Delta T=\Delta T_{max}-T_0\frac{g\cdot p^2}{2mK c_x} \;,
\label{eq:ent2.4}
\end{equation}
 i.e., the temperature variations are proportional to the variations of the kinetic energy. When eq. (\ref{eq:ent2.3}) is inserted into eq. (\ref{eq:ent2.2}), we find

\begin{equation}
S=-\frac{g}{K}\left(\frac{1}{2}Kx^2+\frac{p^2}{2m} \right)=\mbox{const.} \;,
\label{eq:ent2.5}
\end{equation}
where $g/K$ has units of {[1\slash K]}. For a piston coupled to two containers filled with an ideal gas (see Fig. \ref{figure_a1}) as described in \cite{Heimburg2017}, $g/K=3/5T_0$ (see App. \ref{calculatingkgfortheidealgas}). Thus, $E\equiv -\frac{5}{3}T_0\cdot S$, and the equations of motion can be derived from both the energy $E$ or the entropy $S$. The temperature changes of the spring consisting of two coupled gas pistons can be calculated because the equations of state for adiabatic compression is known for the ideal gas. If we assume that the system is closed, we find temperature oscillations. Dissipated energy leads to heating of the two gas containers. Fig. \ref{figure_00} shows the temperature oscillations for parameters given in the figure legend. $\Delta T$ is defined as the mean temperature of the two containers. .

\section{Oscillations of springs}
\label{oscillationsofsprings}

In the following we consider some simple system such that the analogy with Onsager's phenomenological equations becomes more obvious.


\subsection{Damped oscillation of a spring attached to a mass}
\label{dampedoscillationofaspringattachedtoamass}

Hamilton' s equations of motion are given by $\partial\mathcal{H}/\partial x=-\dot p$ and $\partial\mathcal{H}/\partial p=+ \dot x$. For a spring, $\mathcal{H}=\frac{1}{2}Kx^2+\frac{1}{2m}p^2$. These equations can be written in a form resembling that of Onsager's phenomenological equations:

\begin{equation}\label{eq:theor1.1}  
	\underbrace{\left(\begin{array}{c}             
	\dot{x}	\\
	\dot{p}	
	\end{array}\right)}_{\underline{J}}=
	-
\underbrace{\left(\begin{array}{cc}  
	0 & -1 \\
	+1 & 0 	
	\end{array}\right)}_{\underline{\underline{L}}^A}
\underbrace{\left(\begin{array}{cc}       
	K & 0 \\
	0 & \frac{1}{m} 	
	\end{array}\right)}_{\underline{\underline{g}}}
\underbrace{\left(\begin{array}{c}             
	x	\\
	p	
	\end{array}\right)}_{\underline{\alpha}}
\end{equation}
or
\begin{equation}\label{eq:theor1.2}
\underline{J}=-\underline{\underline{L}}^A\cdot \underline{\underline{g}}\cdot \underline{\alpha}\equiv \underline{\underline{L}}^A\cdot \underline{X}
\end{equation}
where $\underline{J}=\underline{\dot{\alpha}}$ is a vector representing the temporal change of the two variables, $\underline{\underline{L}}^A$ is an antisymmetric unity matrix reflecting the antisymmetric structure of Hamilton' s equations of motion, $\underline{\underline{g}}$ is a coupling matrix containing the constants of the spring system (spring constant and mass). The above systems leads to

\begin{equation}
\label{eq:theor1.3}
p=m\dot{x} \qquad \mbox{and} \qquad m\ddot x+Kx=0 \;.
\end{equation}

The motion of a damped spring with a friction coefficient $\eta$ (see Fig. \ref{figure_01}) can be written as
\begin{equation}\label{eq:theor1.4}  
	\medmath{
\underbrace{\left(\begin{array}{c}             
	\dot{x}	\\
	\dot{p}	
	\end{array}\right)}_{\underline{J}}=
	-\bigg[\underbrace{\left(\begin{array}{cc} 
	0 & 0 \\
	0 & \eta 	
	\end{array}\right)}_{\underline{\underline{L}}^S}+
\underbrace{\left(\begin{array}{cc}  
	0 & -1 \\
	+1 & 0 	
	\end{array}\right)}_{\underline{\underline{L}}^A}\bigg]\underbrace{\left(\begin{array}{cc}		       
K & 0 \\
	0 & \frac{1}{m} 	
	\end{array}\right)}_{\underline{\underline{g}}}
\underbrace{\left(\begin{array}{c}             
	x	\\
	p	
	\end{array}\right)}_{\underline{\alpha}}
	}
\end{equation}

or
\begin{equation}\label{eq:theor1.5}
\underline{J}=-\left[\underline{\underline{L}}^S+\underline{\underline{L}}^A\right]\cdot \underline{\underline{g}}\cdot \underline{\alpha} \;,
\end{equation}
where $\underline{\underline{L}}^S$ is a symmetric matrix containing friction. Eq. (\ref{eq:theor1.5}) is consistent with the differential equation
\begin{equation}\label{eq:theor1.6}
m\ddot{x}+\eta\dot{x}+Kx=0 \;.
\end{equation}

The eigenvalues of eq. (\ref{eq:theor1.4}) are\\
\begin{equation}
	\label{eq:theor2.1}
	\lambda_{1,2}=\frac{-\eta\pm\sqrt{\eta^2-4mK}}{2m}
\end{equation}
In the limit of $\eta\rightarrow 0$ the eigenvalues $\lambda_{1,2}=\pm i\sqrt{K/m}\pm i \omega_0$ are the two eigenvalues of a harmonic oscillator that are purely imaginary. In the limit of small $m$ , the eigenvalues are $\lambda_1=-K/\eta$ and $\lambda_2\rightarrow-\infty$. If $\eta^2>4mK$, one finds two real negative eigenvalues. If $\eta^2<4mK$, one finds two conjugate complex eigenvalues.
The time evolution of the positions of the masses and of the momenta is given by

	\begin{equation}\thickmuskip=0mu
\label{eq:theor2.3}
\underline{\alpha}(t)=A_1\cdot \underline{v}_1\exp(\lambda_1 t)+A_2\cdot \underline{v}_2\exp(\lambda_2 t) \;,
\end{equation}
where $\lambda_1$ and $\lambda_2$ are the two eigenvalues and $v_1$ and $v_2$ are the corresponding eigenvectors. The $A_i$ have units of {[kg $\cdot$ m\slash s]}.
In the limit of $m\rightarrow 0$, one can find initial conditions where the first term on the right hand side disappears because the eigenvalue $\lambda_1\rightarrow -\infty$. Thus,

\begin{equation}\thickmuskip=0mu
\label{eq:theor2.4}
\underline{\alpha}(t)=A_2\cdot \underbrace{\left(\begin{array}{c}
1/\eta\\
1
\end{array}\right)}_{v_2}\exp\left(-\frac{K}{\eta} t\right)
\end{equation}

Hence, eq. (\ref{eq:theor2.4}) leads to

\begin{equation}\thickmuskip=0mu
\label{eq:theor2.5}
\dot{x}=-\frac{K}{\eta}x\equiv LX \;,
\end{equation}

with $L={1/\eta}$ and $X=-Kx$.

Summarizing, one generally finds two eigenvalues. Only in the limit of vanishing inertia one of them can be omitted for suitable initial conditions. This is the case described by Onsager.

In non-equilibrium thermodynamics, the forces are defined isothermally and not adiabatically as in analytical mechanics. To take this into account, one can rewrite the above equations. We find from section \ref{phenomenologicalequations} letting $g_{11}\equiv g$, $g_{22}=\frac{K}{g\cdot m}$ and $L_{12}^A=-1$:
\begin{equation}\label{eq:theor2b.1}  
\medmath{
\left(\begin{array}{c}             
	\dot{\alpha}_1	\\
	\dot{\alpha}_2	
	\end{array}\right)=
	-\bigg[\left(\begin{array}{cc} 
	0 & 0 \\
	0 & \eta' 	
	\end{array}\right)+
\left(\begin{array}{cc}  
	0 & -1 \\
	1 & 0 	
	\end{array}\right)\bigg]\left(\begin{array}{cc}
g & 0 \\
	0 & \frac{K}{g\cdot m} 	
	\end{array}\right)
\left(\begin{array}{c}             
	\alpha_1	\\
	\alpha_2	
	\end{array}\right) \;,
}
\end{equation}
where $\alpha_1=x$, $\alpha_2=\frac{g}{K}\cdot m\dot{x}$ and $\eta'=\frac{g}{K}\cdot \eta$.

In the absence of friction, the entropy is given by

\begin{equation}
	S-S_0=-\frac{1}{2}g x^2-\frac{1}{2}\frac{K}{g\cdot m}\alpha_2^2 \;.
\label{eq:theor2b.2}
\end{equation}
Using eq.(\ref{eq:ent3.9}), the equivalents to Hamilton's equations of motion is

\begin{align}
X_1&=\frac{\partial S}{\partial x}=-gx=\dot{\alpha}_2=\frac{g\cdot m}{K}\ddot{x} \;,\nonumber\\
X_2&=\frac{\partial S}{\partial \alpha_2}=-\frac{K}{g\cdot m}\alpha_2=-\dot{x} \;.
\end{align}
The first equation yields $Kx=-m\ddot{x}$, which is Newton's second law. The eigenfrequency is given by $\omega^2=(L_{12}^A)^2 det(\underline{\underline{g}})=\frac{K}{m}$. In the limit of $m\rightarrow 0$, we obtain $\dot{x}=-\frac{g}{\eta'}x$. This is the simplest of Onsager's equations. Thus, while in the entropy-representation the coordinates have a slightly different meaning and the parameters of the matrices assume different values, Newton's second law and the eigenfrequency result unchanged. This is not surprising since entropy and energy are closely related as shown in eq. (\ref{eq:ent2.5}). Note that $g/K\propto 1/T$.

The thermodynamic force in non-equilibrium thermodynamics is derived from an isothermal compression. The observation is that elementary mechanics makes the tacit assumption that K describes isentropic rather than isothermal compression. Thus, the forces in nonequilibrium thermodynamics and in elementary mechanics are defined differently. This is the origin of the formal differences in the equations (\ref{eq:theor1.4}) and (\ref{eq:theor2b.1}) which otherwise are identical, and therefore lead to the same oscillations.

Temperature changes in the damped spring will behave qualitatively similar to those of the gas pistons shown in Fig. \ref{figure_00}, but to obtain a value of $g/K$ one needs to know the equation of state of the spring, which can be obtained experimentally. Further, one has to consider that the dissipation of the total energy of initial compression will lead to a temperature increase of the equilibrated system if the spring is not in contact with any other medium. The calculation of the temperature change requires the knowledge of the heat capacity of the spring.


\subsection{Two masses with three springs}
\label{twomasseswiththreesprings}

\begin{figure}[htbp]
\centering
\includegraphics[width=225pt,height=51pt]{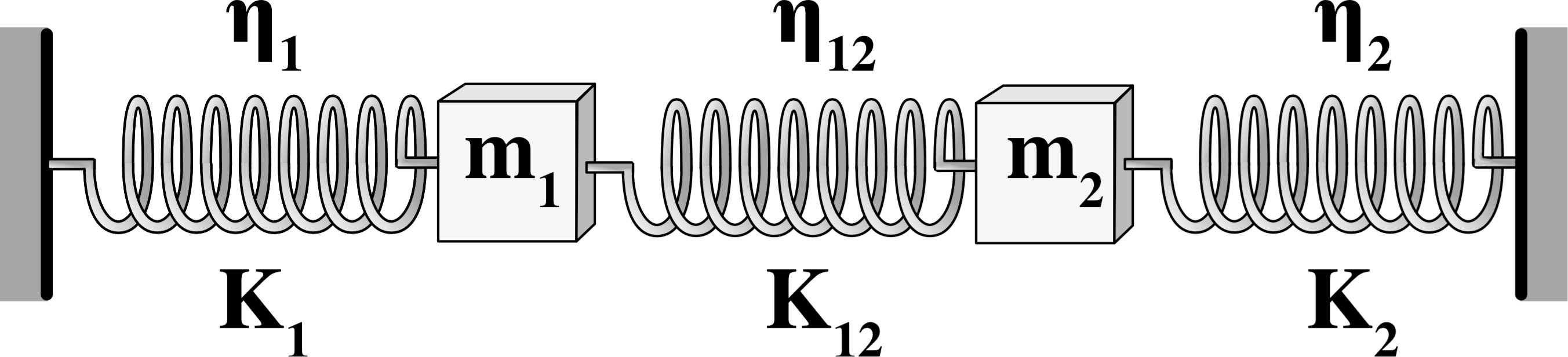}
\caption{Two masses $m_1$ and $m_2$ connected by three springs with spring constants $K_1$, $K_{12}$ and $K_2$. Dissipation occurs in the springs with friction coefficients $\eta_1$, $\eta_{12}$ and $\eta_2$.}
\label{figure_02b}
\end{figure}

Let us consider the case of two masses, $m_1$ and $m_2$ coupled by three springs with spring constants $K_1$, $K_{12}$ and $K_2$ as shown in Fig. \ref{figure_02b}. For simplification, let us set $m_1=m_2\equiv m$ and $K_1=K_{12}=K_2\equiv K$.

The equations of motion can be written as

	\begin{equation}\thickmuskip=0mu
\medmath{
\label{eq:theor3.1}
\underbrace{\left(\!\begin{array}{c}
\dot{x}_1	\\
\dot{x}_2	\\
\dot{p}_1	\\
\dot{p}_2 
\end{array}\!\!\right)}_{\underline{J}}=
-\underbrace{\left(\!\begin{array}{cccc}
0 & 0 & -1 & 0 	\\
0 & 0 & 0 & -1 	\\
1 & 0 & 0 & 0 	\\
0 & 1 & 0 & 0 	
\end{array}\right)}_{\underline{\underline{L}}^A}
\underbrace{\left(\begin{array}{cccc}
2K & -K & 0 & 0 	\\
-K &  2K & 0 & 0	\\
0 & 0 & \frac{1}{m} & 0 	\\
0 & 0 & 0 & \frac{1}{m} 
\end{array}\right)}_{\underline{\underline{g}}}
\underbrace{\left(\begin{array}{c}
x_1	\\
x_2	\\
p_1	\\
p_2	
\end{array}\right)}_{\underline{\alpha}}
}
\end{equation}

In the presence of viscous friction in the three springs (with $\eta_1=\eta_{12}=\eta_2\equiv \eta$, see Fig. \ref{figure_02b}), the equations of motion can be written as
\begin{equation}
	\underline{J}=-\left[\underline{\underline{L}}^S+\underline{\underline{L}}^A\right]\underline{\underline{g}}\;\underline{\alpha}
	\label{eq:theor3.2}
\end{equation}
with
\begin{equation}
\medmath{
\label{eq:theor3.3}
\underline{\underline{L}}^S=\left(\begin{array}{cccc}
0 & 0 & 0 & 0 	\\
0 & 0 & 0 & 0 	\\
0 & 0 & 2\eta & -\eta 	\\
0 & 0 & -\eta & 2\eta	
\end{array}\right)
}\;.
\end{equation}

\noindent Eq. (\ref{eq:theor3.2}) leads to

\begin{align}
\label{eq:theor3.4}
m\ddot{x}_1+\eta(2\dot{x}_1-\dot{x}_2) + K(2x_1-x_2)=0 \;,\nonumber\\
m\ddot{x}_2+\eta(2\dot{x}_2-\dot{x}_1) + K(2x_2-x_1)=0 \;.
\end{align}
The eigenvalues are given by

\begin{align}
	\label{eq:theor3.5}
	\lambda_{1,2}^1&=\frac{-\eta\pm\sqrt{\eta^2-4mK}}{2m}\nonumber\\
	\lambda_{1,2}^2&=\frac{-3\eta\pm3\sqrt{\eta^2-\frac{4}{3} m K}}{2m}
\end{align}

For $m\rightarrow 0$, two eigenvalues in eq. (\ref{eq:theor3.5}) approach $-\infty$ (and can therefore be ignored for suitable initial conditions), and the other two approach $-K/\eta$ and $-3K/\eta$. For $\eta\rightarrow 0$, one finds $\lambda_{1,2}^1=\pm i\sqrt{K/m}=\pm i\omega_0$ and $\lambda_{1,2}^2=\pm i\sqrt{3K/m}$ $=\pm i\sqrt{3}\omega_0$. Under all other circumstances one obtains four finite real or complex eigenvalues that depend on the inertial mass and the friction in the system.

Since in the limit of $m\rightarrow 0$ only two eigenvalues are different from $-\infty$, one can condense eq. (\ref{eq:theor3.5}) into an equation with only two variables and quadratic matrices. We obtain
\begin{equation}\label{eq:theor3.7}  
\underbrace{\left(\begin{array}{c}             
	\dot{x_1}	\\
	\dot{x_2}	
	\end{array}\right)}_{\underline{J}}=
	-\underbrace{\left(\begin{array}{cc}  
	\frac{2}{3\eta} & \frac{1}{3\eta} \\
	\frac{1}{3\eta} & \frac{2}{3\eta} 
	\end{array}\right)}_{\underline{\underline{L}}^{S}}
\underbrace{\left(\begin{array}{cc}       
	2K & -K \\
	-K & 2K 	
	\end{array}\right)}_{\underline{\underline{g}}}
\underbrace{\left(\begin{array}{c}             
	x_1	\\
	x_2	
	\end{array}\right)}_{\underline{\alpha}} \;,
\end{equation}
which displays the same finite eigenvalues as eq. (\ref{eq:theor3.5}) in the limit of $m\rightarrow 0$. Using the thermodynamic forces instead, we find
\begin{equation}\label{eq:theor3.7b}  
\underbrace{\left(\begin{array}{c}             
	\dot{x_1}	\\
	\dot{x_2}	
	\end{array}\right)}_{\underline{J}}=
	-\underbrace{\left(\begin{array}{cc}  
	\frac{2}{3\eta'} & \frac{1}{3\eta'} \\
	\frac{1}{3\eta'} & \frac{2}{3\eta'} 
	\end{array}\right)}_{\underline{\underline{L}}^{S}}
\underbrace{\left(\begin{array}{cc}       
	2g & -g \\
	-g & 2g 	
	\end{array}\right)}_{\underline{\underline{g}}}
\underbrace{\left(\begin{array}{c}             
	x_1	\\
	x_2	
	\end{array}\right)}_{\underline{\alpha}} \;,
\end{equation}
\noindent with $\eta'=\frac{g}{K}\cdot \eta$.
This corresponds to Onsager's phenomenological equations. We see that in the limit of vanishing mass, the $4 \times 4$ system reduces to a $2 \times 2$ system where the variables are $x_1$ and $x_2$ are positions, i.e., extensive quantities. The matrix $\underline{\underline{L}}^{S}$ is symmetric and real positive definite. Eq. (\ref{eq:theor3.7b}) has the form $\underline{J}=-\underline{\underline{L}}^S\underline{\underline{g}}\underline{\alpha}\equiv \underline{\underline{L}}^S\underline{X}$, where $\underline{\underline{L}}^S$ is the symmetric coupling matrix of Onsager's phenomenological equations, and $\underline{X}=-\underline{\underline{g}}\underline{\alpha}$ are the thermodynamic forces.

Summarizing, we find that the general eq.(\ref{eq:theor3.2}) has two limiting cases: 1. for $\eta\rightarrow 0$ we obtain Hamiltonian equations of motion as expressed in eq. (\ref{eq:theor3.1}) and 2. for $m\rightarrow 0$ Onsager's phenomenological equations as described by eq. (\ref{eq:theor3.7b}). The temperature of the springs will vary as a function of the positions of the two masses until the equilibrium position is assumed.

In full analogy to the two spring system described above, one can easily write down the set of equations for an arbitrary number of linearly coupled springs. If there are n mass point, there will be n pairs of variables with coordinates $x_i$ and $p_i$, a symmetric matrix $\underline{\underline{L}}^S$ containing friction coefficients, and antisymmetric matrix $\underline{\underline{L}}^A$ reflecting the antisymmetric structure of Hamilton's equations of motion and a matrix $\underline{\underline{g}}$ describing the metric of the entropy potential which is always symmetric.


\section{Simple electrical circuits}
\label{simpleelectricalcircuits}

The above equations can be generalized to other pairs of extensive and intensive variables. In the case of an electrical oscillator, the variables are related to charge and current.

\begin{figure}[htbp]
\centering
\includegraphics[width=141pt,height=109pt]{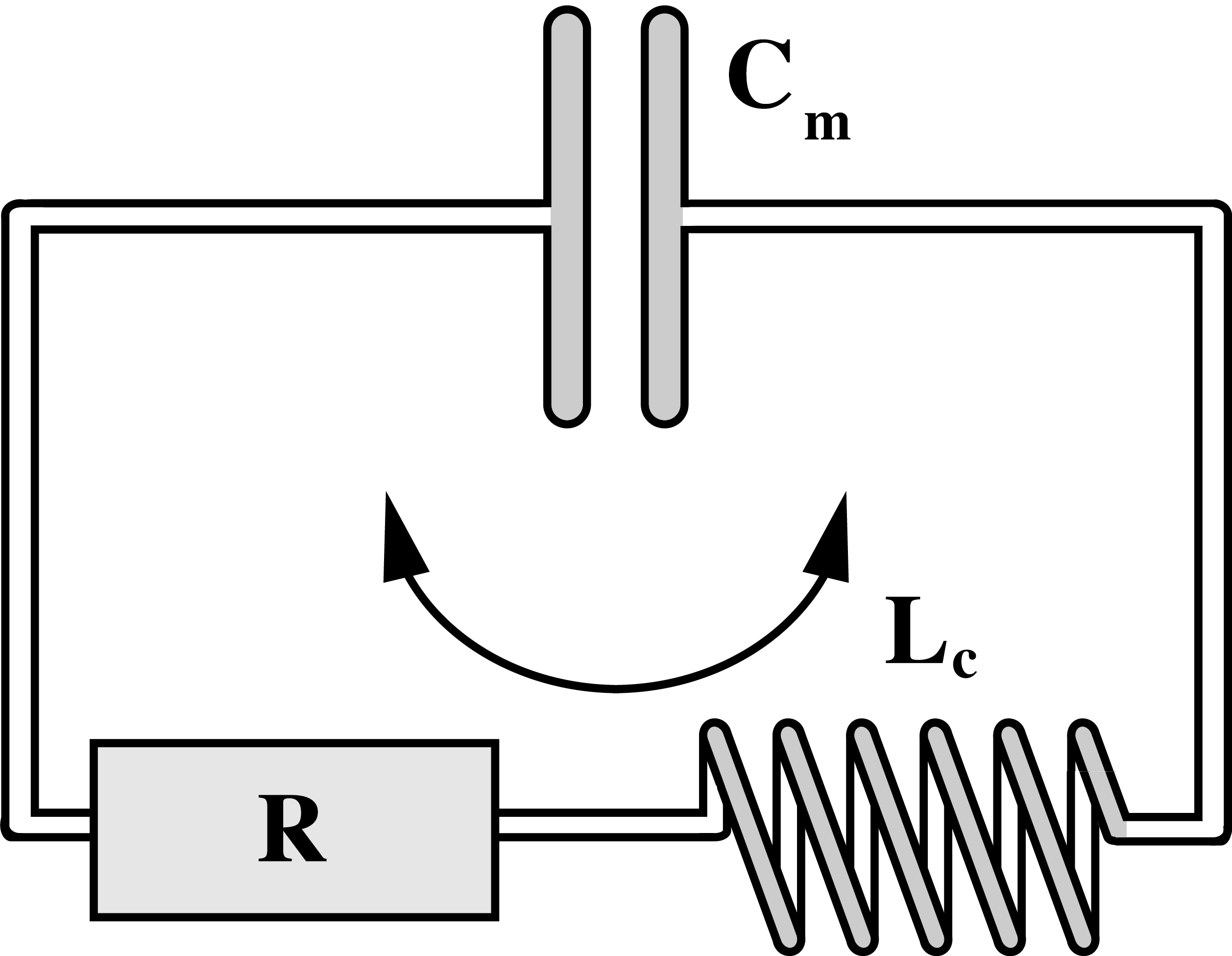}
\caption{Simple electrical circuit with a capacitor with capacitance $C_m$, a resistor with resistance $R$ and a coil with an inductance $L_c$.}
\label{figure_07}
\end{figure}

In analogy to section \ref{dampedoscillationofaspringattachedtoamass} we can describe the damped electrical oscillator shown in Fig. \ref{figure_07} by

\begin{equation}\label{eq:electric1.1}  
\medmath{
\underbrace{\left(\begin{array}{c}             
	\dot{q}	\\
	L_c\dot{I}	
	\end{array}\right)}_{\underline{J}}=
	-\bigg[\underbrace{\left(\begin{array}{cc} 
	0 & 0 \\
	0 & R 	
	\end{array}\right)}_{\underline{\underline{L}}^S}+
\underbrace{\left(\begin{array}{cc}  
	0 & -1 \\
	+1 & 0 	
	\end{array}\right)}_{\underline{\underline{L}}^A}\bigg]\underbrace{\left(\begin{array}{cc}		       
\frac{1}{C_m} & 0 \\
	0 & \frac{1}{L_c} 	
	\end{array}\right)}_{\underline{\underline{g}}}
\underbrace{\left(\begin{array}{c}             
q	\\
L_c I	
	\end{array}\right)}_{\underline{\alpha}}
}
\end{equation}
where $q$ is the charge on the capacitor, $I=\dot{q}$ is the electrical current, $C_m$ is the capacitance, $L_c$ is the inductance and $R$ is the resistance. $\underline{\underline{L}}=(-\underline{\underline{L}}^S+\underline{\underline{L}}^A)\cdot \underline{\underline{g}}$) displays the eigenvalues

\begin{equation}
	\label{eq:electric1.3}
	\lambda_{1,2}=\frac{-R\pm\sqrt{R^2-4\frac{L_c}{C_m}}}{2L_c} \;.
\end{equation}

Eq. (\ref{eq:electric1.1}) is consistent with
\begin{equation}\label{eq:electric1.2}
L_c\ddot{q}+R\dot{q}+\frac{1}{C_m}q=0 \;.
\end{equation}

We have shown in \cite{Heimburg2017} that the equivalent of Newton's second law (in the absence of friction) is given by

\begin{equation}\label{eq:electric1.4b}
L_c\ddot{q}=-\frac{1}{C_m}q
\end{equation}
where the inductance $L_c$ is the equivalent of the inertial mass and $1/C_m$ is the equivalent of the spring constant.

In the limit of $L\rightarrow 0$, the eigenvalues are $\lambda_1=-\infty$ and $\lambda_2=-\frac{1}{R C_m}$ with

\begin{equation}\thickmuskip=0mu
\label{eq:electric1.5}
\dot{q}=-\frac{1}{R C_m}q=\frac{1}{R}\Delta V\equiv L X\;,
\end{equation}
in which $q/C_m=-\Delta V$. This is Ohm's law and corresponds to Onsager's phenomenological equation for one variable, and a phenomenological coefficient $L=1/R$.

In the limit of $R\rightarrow 0$, the eigenvalues are $\lambda_{1,2}=$ $\linebreak$ $ \pm i\sqrt{\frac{1}{L_c C_m}}\equiv \pm i\omega_0$.\\

\textbf{Temperature changes:} The differential of the internal energy of a capacitor is $dE=\Psi dq$ only if the entropy is constant. This allows to calculate the temperature changes of the oscillator. Adiabatic charging of a capacitor implies that for the entropy of the capacitor \cite{Heimburg2017}
\begin{equation}
	dS=\underbrace{\left(\frac{\partial S}{\partial T}\right)_q}_{c_q/T} dT+\left(\frac{\partial S}{\partial q}\right)_T dq=0\;,
	\label{eq:electric1.7}
\end{equation}
where $c_q=T(\partial S/\partial T)_q$ is the heat capacity of the capacitor at constant charge. The Maxwell relation
\begin{equation}
(\partial S/\partial q)_T=-(\partial \Psi/\partial T)_q\equiv g
	\label{eq:electric1.8}
\end{equation}
shows that $(\partial S/\partial q)_T$ can be determined experimentally from the temperature dependence of the voltage on the capacitor at constant charge.

The energy of the undamped circuit is given by
\begin{equation}
E=\frac{1}{2C_m}q^2+\frac{1}{2}L_c I^2=const. 
	\label{eq:electric1.9}
\end{equation}

If we approximate the entropy close to equilibrium by

\begin{equation}
	S=S_0-\frac{1}{2}g q^2 \qquad\qquad T=\mbox{const.}
	\label{eq:electric1.10}
\end{equation}

We find that the temperature change for small $q$ is given by

\begin{equation}
	\Delta T=\frac{T_0}{2c_q}g q^2 \;.
	\label{eq:electric1.11}
\end{equation}

An undamped oscillating LC-circuit will display oscillating temperature changes of the dielectric in the capacitor \cite{Schiffer2006}. The temperature change upon charging a capacitor is known as the electro-caloric effect \cite{Mischenko2006, Scott2011, Crossley2016, Janssen2017}.

The entropy is now given by
\begin{equation}
S=-\frac{1}{2}gq^2+\frac{c_q}{T_0}\Delta T=const.
	\label{eq:electric1.12}
\end{equation}
By comparing eqs. (\ref{eq:electric1.12}) with (\ref{eq:electric1.9}) we find that
\begin{equation}
S=-g\cdot C_m\left(\frac{1}{2C_m}q^2+\frac{1}{2}L_c I^2\right)=const. \;,
	\label{eq:electric1.13}
\end{equation}
or $S=-g C_m\cdot E$. The equivalent of the factor $\frac{g}{K}$ for the oscillating spring is here $g\cdot C_m$, which also has units of $[1/K]$.

We can rewrite eq. (\ref{eq:electric1.1}) by using the thermodynamics forces. We find from section \ref{phenomenologicalequations} letting $g_{11}\equiv g$, $g_{22}=1/g C_m L_c)$ and $L_{12}^A=-1$ that eq. (\ref{eq:electric1.1}) turns into
\begin{eqnarray}\label{eq:electric1.13}  
\bigg(\begin{array}{c}             
	\dot{q}	\\
	g C_m L_c\ddot{q}	
	\end{array}\bigg)&=&
	-\bigg[\bigg(\begin{array}{cc} 
	0 & 0 \\
	0 & R' 	
	\end{array}\bigg)+\bigg(\begin{array}{cc}  
	0 & -1 \\
	1 & 0 	
	\end{array}\bigg)\bigg]\times\nonumber\\
	&&\bigg(\begin{array}{cc}
g & 0 \\
	0 & \frac{1}{gC_mL} 	
	\end{array}\bigg)
\bigg(\begin{array}{c}             
	q\\
	g C_m L_c \dot{q}	
	\end{array}\bigg)
\end{eqnarray}
where $R'=gC_m\cdot R$. The difference between eq. (\ref{eq:electric1.1}) and eq. (\ref{eq:electric1.13}) is only the different definition of the force $X_1=-gq$, where $g$ is an isothermal modulus, and $X_2=-g C_m L_c \dot{q}$.


\section{Chemical oscillations}
\label{chemicaloscillations}

In the following we show that close to stable fix points chemical oscillations can be written in a manner analogous to the oscillations of a spring or an electrical LC-circuit.


\subsection{The Lotka-Volterra scheme}
\label{thelotka-volterrascheme}

Let us consider a simple chemical oscillation from a substrate $S$ to a product $P$ inspired by the Lotka-Volterra reaction:

\begingroup
\begin{align}
\label{eq:chem1.1}
	S+x & \stackrel{k_1}{\rightarrow} 2x\nonumber\\
	x+y & \stackrel{k_2}{\rightarrow} 2y\\
	y & \stackrel{k_3}{\rightarrow}  P\nonumber\\
	\rule{2cm}{1pt} & \rule{2cm}{1pt}\nonumber\\
	S&\rightarrow P\nonumber
\end{align}
\endgroup

This reaction has two intermediates, $x$ and $y$ that may oscillate.

The rate equations of intermediates obey the following non-linear rate equations

\begin{align}
\label{eq:chem1.2}
\dot{[x]}=&(k_1[S]-k_2 [y]) [x] \;,\nonumber\\
	\dot{[y]}=&(k_2 [x]-k_3) [y] \;,
\end{align}
where $k_1$, $k_2$ and $k_3$ are rate constants, $[S]$ is the substrate concentration, and $[x]\equiv x$ and $[y]\equiv y$ (in the following we omit the brackets) the concentrations of the intermediates, respectively. These equations basically follow mass-action kinetics. This equation system possesses one non-zero stable fix-point

\begin{equation}
\label{eq:chem1.3}
\left( \begin{array}{c}
x_0 \\
y_0
\end{array}
 \right)=
 \left( \begin{array}{c}
\frac{k_3}{k_2} \\
\frac{k_1[S]}{k_2}
\end{array}
 \right) \;,
 \end{equation}
for which $\dot{x}=0$ and $\dot{y}=0$. Now one can expand $\dot{x}\equiv f(x,y)$ and $\dot{y}\equiv g(x,y)$ around the fixed point:

\begin{align}
\label{eq:chem1.4}
\dot{x}\equiv & f(x,y)=f(x_0, y_0)+\left(\frac{\partial f}{\partial x}\right)_0(x-x_0)\nonumber\\
&+\left(\frac{\partial f}{\partial y}\right)_0(y-y_0)+...\nonumber\\
	\dot{y}\equiv & g(x,y)=g(x_0, y_0)+\left(\frac{\partial g}{\partial x}\right)_0(x-x_0)\nonumber\\
	&+\left(\frac{\partial g}{\partial y}\right)_0(y-y_0)+...
\end{align}

When we let $u=(x-x_0)$ and $v=(y-y_0)$, which are the deviations of $x$ and $y$ from the stable point, we can rewrite this as

\begin{equation}
\label{eq:chem1.5}
\left( \begin{array}{c}
\dot{u} \\
\dot{v}
\end{array}
 \right)=\underbrace{
 \left( \begin{array}{cc}
(\partial f/\partial x)_0 & (\partial f/\partial y)_0 \\
(\partial g/\partial x)_0 & (\partial g/\partial y)_0
\end{array}
\right)}_{J_0}
\left( \begin{array}{c}
u \\
v
\end{array}
 \right)
 \end{equation}
where $J_0$ is the Jacobian matrix. At the stable fix point, the Jacobian is given by

\begin{equation}
\label{eq:chem1.6}
	J_0=
 \left( \begin{array}{cc}
0 & -k_3 \\
k_1[S] & 0
\end{array}
\right)
 \end{equation}
with eigenvalues $\lambda_{1,2}=\pm i\sqrt{k_1[S]k_3}\equiv \pm i\omega$, i.e., we obtain oscillatory solutions.

Another way of writing this is

\begin{equation}
\medmath{
\label{eq:chem1.7}
\left( \begin{array}{c}
\dot{v} \\
\dot{u}
\end{array}
 \right)=-\underbrace{\left( \begin{array}{cc}
0 & -1\\
1 & 0
\end{array}
 \right)}_{\underline{\underline{L}}^A}
 \underbrace{\left( \begin{array}{cc}
k_3 & 0\\
0 & k_1[S]
\end{array}
 \right)}_{\underline{\underline{g}}}
 \left( \begin{array}{c}
v \\
u
\end{array}
 \right) \;,
}
 \end{equation}
which is identical in structure to eq. (\ref{eq:theor1.1}) for the oscillating spring.
 The above implies that position $x$ and momentum $p$ of the spring are conjugated in exactly the same way as variables $v$ and $u$, where

 \begin{equation}
\label{eq:chem1.8}
 	K\equiv k_3 \qquad \mbox{and} \qquad \frac{1}{m}\equiv  k_1[S]
 \end{equation}

For the linearized Lotka-Volterra, the analogy to Newton's 2nd law ($Kx=-m\ddot{x}$) is:

\begin{equation}
\label{eq:chem1.9}
	k_3 v=-\frac{1}{k_1[S]}\ddot{v}.
\end{equation}

and $u=(1/k_1[S]) \dot{v}$ corresponds to $p=m\dot{x}$.

In the presence of friction, the Lotka-Volterra system can be described by

\begin{equation}
\medmath{
\label{eq:chem1.11}
 \left( \begin{array}{c}
\dot{v} \\
\dot{u}
\end{array}
 \right)=-\bigg[\underbrace{\left( \begin{array}{cc}
0 & 0\\
0 & \eta
\end{array}
 \right)}_{\underline{\underline{L}}^S}+
 \underbrace{\left( \begin{array}{cc}
0 & -1\\
1 & 0
\end{array}
 \right)}_{\underline{\underline{L}}^A}\bigg]
 \underbrace{\left( \begin{array}{cc}
k_3 & 0\\
0 & k_1[S]
\end{array}
 \right)}_{\underline{\underline{g}}}
 \left( \begin{array}{c}
v \\
u
\end{array}
 \right) \;,
}
 \end{equation}
In the over-damped case, there exist two eigenvalues: $\lambda_1=-\frac{k_3}{\eta}$ and $\lambda_2=-\infty$.

We see here that the Lotka-Volterra system can be described by an equation system that is similar to that of an oscillating spring. The main element of a physical oscillation is that potential energy is converted into kinetic energy and vice vera in a process the conserves entropy. However, when looking at the reaction scheme eq. (\ref{eq:chem1.1}) we see that the similarity of the equations for Lotka-Volterra and that of an oscillating spring is of purely mathematical nature because $x$ is converted into $y$ but not vice versa. Both, $x$ and $y$ represent chemical reagents, and $x$ does not correspond to a momentum. Thus, the reaction scheme eq. (\ref{eq:chem1.1}) does not represent a truly reversible physical oscillation but is just a mathematical analogy.
However, one can turn the argument around. The fact that eq. (\ref{eq:chem1.7}) describes the oscillations may indicate that there exist a different mechanism by which a chemical oscillations may function. It is possible that the true mechanism of an oscillating reaction is an oscillating physical system with inertia. If that was the case, the constants $k_1$ , $k_2$, and $k_3$ would find a different interpretation. $1/(k_1[S])$ would correspond to an inertial mass, and $k_3$ to a spring constant. In \cite{Heimburg2017} we have therefore proposed that there may exist a chemical inertia in oscillating reactions. Such a physical oscillation would be related to a non-equilibrium system with entropy conservation. As we have argued further up, an oscillation with entropy conservation would lead to temperature oscillations. Therefore, temperature recordings are crucial to discriminate between the different processes.


\subsection{The Brusselator}
\label{thebrusselator}

Field and collaborator \cite{Field1972, Field1974} proposed a reaction scheme for the Belousov-Zhabotinsky reaction. A similar minimalistic scheme is the Brusselator \cite{Nicolis1971, Nicolis1971b, Kondepudi1998} describing the overall reaction $S_1+S_2 \rightarrow P_1+P_2$. The scheme is composed as a sequence of chemical reactions with two intermediates $x$ and $y$:

\begingroup
\begin{align}
	S_1 & \stackrel{k_1}{\rightarrow} x\nonumber\\
	S_2+x & \stackrel{k_2}{\rightarrow} y+P_1\\
	2x + y & \stackrel{k_3}{\rightarrow}  3x\nonumber\\
	x & \stackrel{k_4}{\rightarrow}  P_2\nonumber\\
	\rule{2cm}{1pt} & \rule{2cm}{1pt}\nonumber\\
	S_1+S_2&\rightarrow P_1+P_2\nonumber
\end{align}
\endgroup

The time dependence of the intermediate concentrations is described by the following coupled non-linear differential equations:

\begin{align}
	\dot{[x]}=&k_1 [S_1]-k_2[S_2] [x]+k_3 [x]^2[y]-k_4[x]\nonumber\\
	\dot{[y]}=&k_2[S_2][x]-k_3[x]^2[y] \;,
\end{align}
where $[S_1]$, $[S_2]$, $[x]\equiv x$ and $[y]\equiv y$ (we again omit the brackets for the intermediates) are concentrations, and the $k_i$ are rate constants. The stable non-zero fix-point of this system is given by

\begin{equation}
 \left( \begin{array}{c}
x_0 \\
y_0
\end{array}
 \right)=
 \left( \begin{array}{c}
\frac{k_1[S_1]}{k_4} \\
\frac{k_2 k_4}{k_1 k_3}\frac{[S_2]}{[S_1]}
\end{array}
 \right) \;.
 \end{equation}
At the fix point, the Jacobian is given by (\cite{Kondepudi1998})

\begin{equation}
	J_0=
 \left( \begin{array}{cc}
k_2[S_2]-k_4 & +\frac{k_1^2 k_3 [S_1]^2}{k_4^2} \\
-k_2 [S_2] & -\frac{k_1^2 k_3 [S_1]^2}{k_4^2}
\end{array}
\right) \;.
 \end{equation}
Using $u=(x-x_0)$ and $v=(y-y_0)$, the linearized equations take the form:

  \begin{equation}
\medmath{
 \left( \begin{array}{c}
\dot{u} \\
\dot{v}
\end{array}
 \right)=\underline{\underline{J}}_0
 \left( \begin{array}{c}
u \\
v
\end{array}
 \right) \;,
}
 \end{equation}
which can also be written as

 \begin{equation}
\medmath{
 \left( \begin{array}{c}
\dot{u} \\
\dot{v}
\end{array}
 \right)=- \bigg[
 \underbrace{\left( \begin{array}{cc}
0 & 0\\
0& \eta
\end{array}
 \right)}_{\underline{\underline{L}}^S}+
 \underbrace{\left( \begin{array}{cc}
0 & -1\\
1 & 0
\end{array}
 \right)}_{\underline{\underline{L}}^A} \bigg]
 \underbrace{\left( \begin{array}{cc}
a & b\\
b & c
\end{array}
 \right)}_{\underline{\underline{g}}}
 \left( \begin{array}{c}
u \\
v
\end{array}
 \right) \;,}
 \end{equation}
where

 \begin{align}
 	b &=k_2[S_2]-k_4 \;,\nonumber\\
 	c &= \frac{k_1^2k_3[S_1]^2}{k_4^2} \;,\\
 	a &= \frac{b^2}{c}+k_4 \;,\nonumber\\
 	\eta &= \frac{c-b}{c} \;.
 \end{align}
The matrix $\underline{\underline{g}}$ is symmetric and positive definite, because $a>0$ and $c>0$. For the undamped system, $\eta=0$ and $c=b$. This happens when

\begin{equation}
	[S_2]=\frac{k_1^2 k_3 [S_1^2]}{k_4^2 k_2}+\frac{k_4}{k_2} \;.
\end{equation}
This is the instability criterium given for the Brusselator by Kondepudi \& Prigogine \cite{Kondepudi1998}, which demonstrates that $\eta$ in fact plays the role of a friction coefficient in the Brusselator. The undamped case is given by
 \begin{equation}
\medmath{
 \left( \begin{array}{c}
\dot{u} \\
\dot{v}
\end{array}
 \right)=
 -\underbrace{\left( \begin{array}{cc}
0 & -1\\
1 & 0
\end{array}
 \right)}_{\underline{\underline{L}}^A}
 \underbrace{\left( \begin{array}{cc}
a & c\\
c & c
\end{array}
 \right)}_{\underline{\underline{g}}}
 \left( \begin{array}{c}
u \\
v
\end{array}
 \right) \;.
}
 \end{equation}
This system has eigenvalues $\lambda_{1,2}=\pm i\sqrt{k_4 c}$ which are imaginary because both $c$ and $k_4$ are positive. Thus, the Brusselator can just as the Lotka-Volterra scheme be compared to a spring-like system, if one finds temperature oscillations in experiments. As shown in \cite{Franck1971, Franck1978, Boeckmann1996, Heimburg2017}, the Belousov-Zhabotinsky reaction in fact displays temperature oscillations. It is interesting to note that the instability condition for the Brusselator corresponds to the case of the undamped oscillation.


\section{Discussion}
\label{discussion}

We consider here the striking experimental observation that mechanical springs \cite{Heimburg2021}, electrical oscillators \cite{Schiffer2006}, oscillating chemical reactions \cite{Franck1971, Franck1978, Boeckmann1996} and oscillating yeast cell preparations \cite{Teusink1996, Thoke2015, Thoke2018} all display oscillations in temperature.
We show that in an undamped physical oscillator both energy and entropy are conserved. For an insulated damped spring this leads to temperature oscillations and ultimately to an increase of the total temperature of the spring. Similar statements can be made for oscillating electrical systems, where the temperature of the capacitor will oscillate.

One can express the equations of motion for springs and other physical oscillator in a manner that makes it easier to see the striking similarities with Onsager's phenomenological equations with the difference that they contain momenta. In the limit of small inertia one finds Onsager's equations in which the thermodynamic fluxes are proportional to the thermodynamic forces, and the matrix generating the couplings is symmetric. This is known as Onsager's reciprocal relations. The underlying assumption is that thermal fluctuations do not display any preferred direction in time. This corresponds to over-damped systems. Forces are defined isothermally from the fluctuations of the extensive variables. In the undamped physical oscillator, the coupling matrix is antisymmetric. This generates the antisymmetry of Hamilton's equations of motion. Forces are defined from isentropic changes in state. Since both, energy and entropy are constant, one can rewrite the equations such that isothermal (thermodynamic) forces are at the center. Undamped oscillators and rate equations as in Onsager's equations are now just the two extreme limits where either friction or inertia approach zero. In all real cases, both inertia and friction are always present. These cases are not described by Onsager's phenomenological equations.

When considering oscillators as thermodynamic systems, one can see that the above concept applies to all pairs of intensive and conjugated extensive variables, not only to pressure and volume, or force and position, respectively. We can therefore also apply it to pairs such as charge and voltage, and in chemistry to a reaction coordinate and the chemical affinity \cite{Heimburg2017}. Electrical systems contain capacitors, inductive coils and resistors. Here, the variables are the charge on the capacitor and the electrical current. One finds analogies to Hamilton's equations of motions and to Newton's second law. The inductance is analogous to an inertial mass \cite{Heimburg2017}. One finds temperature oscillations in the dielectric of the capacitor \cite{Schiffer2006}. This makes clear that inertial terms can be found in systems without moving masses. In the present paper, we attempt to extend this concept to chemical oscillations represented by well-known schemes for oscillating reactions as are the Lotka-Volterra equations and the Brusselator. It has been shown by others that the Lotka-Volterra oscillations display a Hamiltonian character \cite{Kerner1957, Kerner1959, Kerner1964, Nutku1990, Plank1995} and here we make use of this fact. Chemical oscillations such as the Belousov-Zhabotinsky reaction that are described by reaction schemes similar to the Brusselator in fact display temperature oscillations (discussed in \cite{Heimburg2017}). It is therefore tempting and relevant to compare physical with chemical oscillators.

The central element of a mechanical clock is the loaded main spring. It's energy is slowly released via an oscillating balance wheel connected to a spring. Its eigenfrequency determines how fast the main spring is unloaded. The balance wheel is connected to the main spring mechanism via an escapement. In classical pocket watches, the watch is wound and then runs for about a day while slowly dissipating the energy stored in the main spring. Thus, one finds two separate processes in the watch: the unloading of the main spring which represents an overall dissipative process, and a nearly reversible oscillation of the balance wheel with only minor friction. The balance wheel oscillates between states of maximum potential and maximum velocity, which can be seen as two intermediate states of the spring. After the main spring is unloaded, all its energy is dissipated in form of heat. The main purpose of the main spring is to compensate for the friction in the oscillation of the balance wheel. Thus, apart from some tricky details related to the exact functioning of the escapement, the non-equilibrium problem that we treat is basically that of a damped oscillating spring treated in section \ref{dampedoscillationofaspringattachedtoamass} driven by a constant force provided by the main spring.

A chemical oscillation carries some similarities to this process. There exists a substrate $S$ (corresponding to the loaded main spring), and a product $P$ (corresponding to the heat dissipated by the unloaded spring). The chemical reaction may have intermediate products that may display oscillations in concentration (in analogy to the two states of the oscillating balance wheel). This is referred to as a chemical clock. The experimental finding of temperature oscillations suggests the possibility that there exists a reversible, entropy-conserving process that is an elementary part of the oscillating system.

The Lotka-Volterra predator-prey scheme was originally designed to describe oscillating reactions. It consists of a reaction $S\rightarrow P$ with two oscillating intermediates $x$ and $y$. The Belousov-Zhabotinsky reaction is somewhat more complicated. The Field-K{\"o}r{\"o}s-Noyes scheme \cite{Field1972, Field1974} is an attempt to describe it. The so-called Brusselator is a similar minimalistic reaction scheme with two substrates and two products ($S_1+S_2\rightarrow P_2+P_2$) that also displays two oscillating intermediates $x$ and $y$. In analogy with the force of the main spring of a clock, the concentrations of substrates are kept constant when treating the oscillations in both Lotka-Volterra and Brusselator. These reaction schemes are considered far-from-equilibrium processes. Unfortunately, they do not make use of the advanced thermodynamics derived by Prigogine and collaborators containing thermodynamic forces of various nature (e.g., temperature gradients) but represent rather mass action law kinetics that only uses concentrations as variables. They do not consider temperature variations.

We have shown here (and other have shown that before us \cite{Kerner1957, Kerner1959, Kerner1964, Nutku1990, Plank1995}) that the Lotka-Volterra scheme can be written in a manner that resembles that of a spring and the equations of motion resemble that of a Hamiltonian system. We further showed that the equations can be written analogous to Onsager's phenomenological equations but with antisymmetric coupling matrices. This also holds true for the Brusselator. However, as we have already argued in the introduction, this is a mathematical similarity to a Hamiltonian system but not a physical one. The Lotka-Volterra equations and the Brusselator consist only of forward reactions, and there is no reversible conversion from intermediate $x$ to intermediate $y$. In contrast, an oscillating spring displays a reversible conversion of position to momentum. Thus, this setting displays completely different physics but the same mathematics. When it comes to mathematics only, the Lotka-Volterra scheme could be used to describe the oscillations of a mechanical spring as well, when the intermediate $x$ is likened to momentum and the intermediate $y$ is the position. This would clearly represent wrong physics because the underlying reactions are directed. Vice versa, the predator-prey scheme can be described by Hamilton's equations of motion if one renames the predator and prey as position and momentum. This would provide the correct mathematics but would also be inappropriate because predators eat prey but prey doesn't eat predators (they eat gras only) and there is no reversible conversion between the two species. The predator-prey case is not a Hamiltonian system.

How is it, however, with oscillating chemical reactions? Here it is less obvious that they might not be represented by Hamiltonian physics. It is clear that close to a fix-point the oscillations are successfully described by Lotka-Volterra or by the Brusselator. But how can we prove that this is not only a mathematical description of a system that in reality is rather of Hamiltonian nature? The main feature of a Hamiltonian system is the presence of inertia which is absent in Lotka-Volterra and the Brusselator. However, those schemes obviously contain couplings between the rate equations that mimic the presence of inertia and thereby generate similar mathematics.

As we have shown above, the oscillation of a spring conserves entropy. An unavoidable consequence are temperature oscillations of the spring \cite{Heimburg2017}. The temperature oscillations in a spring is immediately obvious when considering pistons filled by ideal gases (see appendix) because the equation of state is well-known. While it is not generally acknowledged, the same is true for metal springs. The equation of state for a metal spring can be measured, and it is unavoidable that one finds temperature oscillations \cite{Heimburg2021}. Likewise, the oscillations of an electrical circuit lead to oscillations of the temperature of the dielectric core of the capacitor. This effect is called the electro-caloric effect \cite{Mischenko2006, Scott2011, Crossley2016, Janssen2017}. Thus, the criterium for equating an oscillation with a Hamiltonian system containing inertia is the observation of temperature oscillations, which are mandatory in Hamiltonian systems but not contained in coupled rate equations. In fact, such oscillations have been found in the Belousov-Zhabotinsky reaction \cite{Franck1971, Franck1978, Boeckmann1996}, the Briggs-Rauscher oscillation \cite{Wang2017c} and in some oscillations of yeast cells \cite{Teusink1996}. The case of metabolic oscillations of yeast cells is especially interesting because one finds not only oscillations in temperature but also in volume and in water activity \cite{Thoke2015, Thoke2018} indicating that there are several variables in the system that oscillate in phase as it is naturally expected for adiabatic oscillations where all extensive variables will oscillate.

It is interesting to note that the action potential in nerves is also displays an adiabatic nature. It was shown that the temperature of a nerve undergoes a reversible change in phase with voltage changes \cite{Abbott1958, Howarth1968, Howarth1975, Ritchie1985} and mechanical variations \cite{Iwasa1980a, Iwasa1980b, GonzalezPerez2016}. The striking consequences were extensively discussed in \cite{Heimburg2021}. It was proposed that the nerve pulse consists of a solitary electromechanical pulse reminiscent of an sound wave \cite{Heimburg2005c}. Due to the nonlinear nature of the compressibility of nerve membranes one finds a single solitary pulse rather than a periodic wave. But the consequence is the same. The reversible temperature change can be considered as a proof for the adiabatic nature of the nerve pulse.

Since so different systems as springs, electrical circuits, coupled chemical reactions and cell populations (and nerves) may display temperature oscillations in phase with periodic variations of other variables, it is plausible that all these cases represent non-equilibrium processes in which a sub-process conserves entropy. This can even happen in close-to-equili\-brium situations. Thus, rather than stating that the oscillations must be the consequence of far-from-equilibrium processes because there is no place for them in the linear regime of nonequilibrium thermodynamics of over-damped systems, one may conclude that oscillations are possible close to equilibrium if one relaxes the constraint of the reciprocal relations, which immediately results in oscillations. This is support by the fact that one can write down the equations of motion for oscillating reaction in a similar manner as an oscillating spring. This suggests the chemical clocks are in fact similar to physical clocks.

If chemical oscillations were in fact analogous to oscillations of springs as suggested by the finding of temperature oscillations, it should be possible to excite or stimulate such oscillations with periodic external forces, e.g., by superimposing a periodic temperature variation. Close to the resonance frequency, such stimulation would result in larger effects than far away from the resonance frequency. In fact, resonant phenomena in the Belousov-Zhabotinsky reaction forced by periodic light pulses have been observed \cite{Lin2000}. In reaction schemes where the only variables are concentrations, such resonances remain obscure. Thus, it seems that the parallels between adiabatic oscillations and chemical oscillators are much stronger as one would intuitively expect.


\section{Conclusion}
\label{conclusion}

While oscillations of springs can be described by Hamilton's equations of motion, chemical oscillations have been treated by systems of directed chemical reactions that can be described by coupled nonlinear rate equations. Due to conservation of entropy, physical oscillations are accompanied by temperature oscillations. The reaction schemes of the Lotka-Volterra reaction and the Brusselator do not conserve entropy and are not obviously linked to temperature oscillations. They are thought to represent far-from-equilibrium systems. Thus, experiments can discriminate between the two mechanisms by measuring temperature changes. The fact that the Belousov-Zhabotinsky reaction is accompanied by temperature oscillation as are growth cycles of yeast cells suggests that the physics of oscillating reactions may be much closer to that of springs than widely assumed.


\vspace{0.2cm}

\noindent \textbf{Conflicts of interest:} There are no conflicts of interest to declare.


\vspace{0.2cm}

\noindent \textbf{Acknowledgments:}

I thank Prof. Andrew D. Jackson from the Niels Bohr International Academy, Prof. Lars F. Olsen from the University of Southern Denmark and Prof. Matej Daniel from the Technical University of Prague for useful discussions and for a critical reading of the manuscript. Prof. Jackson introduced me to the fix-point analysis of the oscillating reactions and discussed the difference between a purely mathematical and a physical analogy, Prof. Olsen corrected some imprecisions about the FKN-oscillator and the Brusselator, and Prof. Daniel suggested the possibility of stimulating oscillating chemical reactions and the observation of resonant behavior.


\begin{thebibliography}{10}
	
	\bibitem{Zaikin1970}
	Zaikin, A.~N., and A.~M. Zhabotinsky.
	\newblock 1970.
	\newblock Concentration wave propagation in two-dimensional liquid-phase
	self-oscillating system.
	\newblock Nature 225:535--537.
	
	\bibitem{Briggs1973}
	Briggs, T.~S., and W.~C. Rauscher.
	\newblock 1973.
	\newblock Oscillating iodine clock.
	\newblock J.\ Chem.\ Education 50:496--496.
	
	\bibitem{Rensing2001}
	Rensing, L., U.~{Meyer-Grahle}, and P.~Ruoff.
	\newblock 2001.
	\newblock Biological timing and the clock metaphor: {O}scillatory and hourglass
	mechanicms.
	\newblock Chronobiology Int. 18:329--369.
	
	\bibitem{Dutt1993}
	Dutt, A.~K.
	\newblock 1993.
	\newblock Asymptotically stable limit cycles in a model of glycolytic
	oscillations.
	\newblock Chem.\ Phys.\ Lett. 208:139--142.
	
	\bibitem{Teusink1996}
	Teusink, B., C.~Larsson, J.~Diderich, P.~Richard, K.~{van Dam}, L.~Gustafsson,
	and H.~V. Westerhoff.
	\newblock 1996.
	\newblock Synchronized heat flux oscillations in yeast cell populations.
	\newblock J.\ Biol.\ Chem. 271:24442--24448.
	
	\bibitem{Thoke2015}
	Thoke, H.~S., A.~Tobiesen, J.~Brewer, P.~L. Hansen, R.~P. Stock, L.~F. Olsen,
	and L.~A. Bagatolli.
	\newblock 2015.
	\newblock Tight coupling of metabolic oscillations and intracellular water
	dynamics in saccharomyces cerevisiae.
	\newblock PloS One 10:e0117308.
	
	\bibitem{Thoke2018}
	Thoke, H.~S., L.~F. Olsen, L.~Duelund, R.~P. Stock, T.~Heimburg, and L.~A.
	Bagatolli.
	\newblock 2018.
	\newblock Is a constant low-entropy process at the root of glycolytic
	oscillations?
	\newblock J.\ Biol.\ Phys. 44:419--431.
	
	\bibitem{Krinsky1978}
	Krinsky, V.~I.
	\newblock 1978.
	\newblock Mathematical models of cardiac arrhythmias (spiral waves).
	\newblock Pharmacol.\ Ther.\ B 3:539--555.
	
	\bibitem{Lakatta2010}
	Lakatta, E.~G., V.~A. Maltsev, and T.~M. Vinogradova.
	\newblock 2010.
	\newblock A coupled system of intracellular {C}a$^{2+}$ clocks and surface
	membrane voltage clocks controls the timekeeping mechanism of the heart's
	pacemaker.
	\newblock Circ.\ Res. 106:659--673.
	
	\bibitem{Vitaterna2001}
	Vitaterna, M.~H., and J.~S. Takahashi.
	\newblock 2001.
	\newblock Overview of circadian rhythms.
	\newblock Alcohol Res.\ \& Health 25:85--93.
	
	\bibitem{Albrecht2012}
	Albrecht, U.
	\newblock 2012.
	\newblock Timing to perfection: the biology of central and peripheral circadian
	clocks.
	\newblock Neuron 74:246--260.
	
	\bibitem{Lotka1925}
	Lotka, A.~J., 1925.
	\newblock Elements of Physical Biology.
	\newblock Williams and Wilkins Company, Baltimore.
	
	\bibitem{Kerner1957}
	Kerner, E.~H.
	\newblock 1957.
	\newblock A statistical mechanics of interacting biological species.
	\newblock Bull.\ Math.\ Biophys. 19:121--146.
	
	\bibitem{Stenseth1997}
	Stenseth, N.~C., W.~Falck, O.~Bj{\o}rnstad, and C.~Krebs.
	\newblock 1997.
	\newblock Population regulation in snowshoe hare and canadian lynx: Asymmetric
	food web configurations between hare and lynx.
	\newblock Proc.\ Natl.\ Acad.\ Sci.\ USA 94:5147--5152.
	
	\bibitem{Nicolis1971}
	Nicolis, G.
	\newblock 1971.
	\newblock Stability and dissipative structures in open systems far from
	equilibrium.
	\newblock Adv.\ Chem.\ Phys. 19:209--324.
	
	\bibitem{Nicolis1971b}
	Nicolis, G., and I.~Prigogine.
	\newblock 1971.
	\newblock Fluctuations in nonequilibrium systems.
	\newblock Proc.\ Natl.\ Acad.\ Sci.\ USA 68:2102--2107.
	
	\bibitem{Kondepudi1998}
	Kondepudi, D., and I.~Prigogine, 1998.
	\newblock Modern thermodynamics.
	\newblock Wiley, Chichester.
	
	\bibitem{Field1972}
	Field, R.~J., E.~K{\"o}r{\"o}s, and R.~Noyes.
	\newblock 1972.
	\newblock Oscillations in chemical systems. {II.} {T}horough analysis of
	temporal oscillation in the bromate-cerium-malonic acid system.
	\newblock J.\ Am.\ Chem.\ Soc. 94:8649--8664.
	
	\bibitem{Field1974}
	Field, R.~J., and R.~M. Noyes.
	\newblock 1974.
	\newblock Oscillations in chemical systems. {IV}. {L}imit cycle behavior in a
	model of a real chemical reaction.
	\newblock J.\ Am.\ Chem.\ Soc. 60:1877--1884.
	
	\bibitem{Kerner1959}
	Kerner, E.~H.
	\newblock 1959.
	\newblock Further considerations on the statistical mechanics of biological
	associations.
	\newblock Bull.\ Math.\ Biophys. 21:217--255.
	
	\bibitem{Kerner1964}
	Kerner, E.~H.
	\newblock 1964.
	\newblock Dynamical aspects of kinetics.
	\newblock Bull.\ Math.\ Biophys. 26:333--349.
	
	\bibitem{Nutku1990}
	Nutku, Y.
	\newblock 1990.
	\newblock Hamiltonian structure of the {L}otka-{V}olterra equations.
	\newblock Phys.\ Lett.\ A 145:27--28.
	
	\bibitem{Plank1995}
	Plank, M.
	\newblock 1995.
	\newblock Hamiltonian structures for the n--dimensional {L}otka--{V}olterra
	equations.
	\newblock J.\ Math.\ Phys. 36:3520--3534.
	
	\bibitem{Onsager1931b}
	Onsager, L.
	\newblock 1931.
	\newblock Reciprocal relations in irreversible processes. {II.}
	\newblock Phys.\ Rev. 38:2265--2279.
	
	\bibitem{Heimburg2021}
	Heimburg, T.
	\newblock 2021.
	\newblock The important consequences of the reversible heat production in
	nerves and the adiabaticity of the action potential.
	\newblock Progr.\ Biophys.\ Mol.\ Biol. 162:26--40.
	
	\bibitem{Mischenko2006}
	Mischenko, A.~S., Q.~Zhang, J.~F. Scott, R.~W. Whatmore, and N.~D. Mathur.
	\newblock 2006.
	\newblock Giant electrocaloric effect in thin-film
	{P}b{Z}r$_{0.95}${T}i$_{0.05}${O}$_3$.
	\newblock Science 311:1270--1271.
	
	\bibitem{Scott2011}
	Scott, J.~F.
	\newblock 2011.
	\newblock Electrocaloric materials.
	\newblock Annu.\ Rev.\ Mater.\ Res. 41:229--240.
	
	\bibitem{Crossley2016}
	Crossley, S., T.~Usui, B.~Nair, S.~Kar-Narayan, X.~Moya, S.~Hirose, A.~Ando,
	and N.~D. Mathur.
	\newblock 2016.
	\newblock Direct electrocaloric measurement of
	0.9{P}b({M}g$_{1/3}${N}b$_{2/3}$){O}$_3$-0.1{P}b{T}i{O}$_3$ films using
	scanning thermal microscopy.
	\newblock Appl.\ Phys.\ Letters 108:032902.
	
	\bibitem{Janssen2017}
	Janssen, M., and R.~{van Roij}.
	\newblock 2017.
	\newblock Reversible heating in electric double layer capacitors.
	\newblock Phys.\ Rev.\ Lett. 118:096001.
	
	\bibitem{Heimburg2017}
	Heimburg, T.
	\newblock 2017.
	\newblock Linear nonequilibrium thermodynamics of reversible periodic processes
	and chemical oscillations.
	\newblock Phys.\ Chem.\ Chem.\ Phys. 19:17331--17341.
	
	\bibitem{Franck1971}
	Franck, U., and W.~Geiseler.
	\newblock 1971.
	\newblock Zur periodischen {R}eaktion von {M}alons{\"a}ure mit {K}aliumbromat
	in {G}egenwart von {C}er-{I}onen.
	\newblock Naturwissenschaften 58:52--53.
	
	\bibitem{Franck1978}
	Franck, U.~F.
	\newblock 1978.
	\newblock Chemical oscillations.
	\newblock Angew.\ Chem.\ Int.\ Edit. 17:1--15.
	
	\bibitem{Boeckmann1996}
	B\"ockmann, M., B.~Hess, and S.~M\"uller.
	\newblock 1996.
	\newblock Temperature gradients traveling with chemical waves.
	\newblock Phys.\ Rev.\ E 53:5498--5501.
	
	\bibitem{Abbott1958}
	Abbott, B.~C., A.~V. Hill, and J.~V. Howarth.
	\newblock 1958.
	\newblock The positive and negative heat production associated with a nerve
	impulse.
	\newblock Proc.\ Roy.\ Soc.\ Lond.\ B 148:149--187.
	
	\bibitem{Howarth1968}
	Howarth, J.~V., R.~Keynes, and J.~M. Ritchie.
	\newblock 1968.
	\newblock The origin of the initial heat associated with a single impulse in
	mammalian non-myelinated nerve fibres.
	\newblock J.\ Physiol. 194:745--793.
	
	\bibitem{Howarth1975}
	Howarth, J.
	\newblock 1975.
	\newblock Heat production in non-myelinated nerves.
	\newblock Phil. Trans. Royal Soc. Lond. 270:425--432.
	
	\bibitem{Ritchie1985}
	Ritchie, J.~M., and R.~D. Keynes.
	\newblock 1985.
	\newblock The production and absorption of heat associated with electrical
	activity in nerve and electric organ.
	\newblock Quart.\ Rev.\ Biophys. 18:451--476.
	
	\bibitem{Schiffer2006}
	Schiffer, J., D.~Linzen, and D.~U. Sauer.
	\newblock 2006.
	\newblock Heat generation in double layer capacitors.
	\newblock J.\ Power Sources 160:765--772.
	
	\bibitem{Wang2017c}
	Wang, T., 2017.
	\newblock Studies on the Action Potential From a Thermodynamic Perspective.
	\newblock Ph.D. thesis, Faculty of {S}cience, {U}niversity of {C}openhagen.
	
	\bibitem{Iwasa1980a}
	Iwasa, K., and I.~Tasaki.
	\newblock 1980.
	\newblock Mechanical changes in squid giant-axons associated with production of
	action potentials.
	\newblock Biochem.\ Biophys.\ Research Comm. 95:1328--1331.
	
	\bibitem{Iwasa1980b}
	Iwasa, K., I.~Tasaki, and R.~C. Gibbons.
	\newblock 1980.
	\newblock Swelling of nerve fibres associated with action potentials.
	\newblock Science 210:338--339.
	
	\bibitem{GonzalezPerez2016}
	{Gonzalez-Perez}, A., L.~D. Mosgaard, R.~Budvytyte, E.~{Villagran Vargas},
	A.~D. Jackson, and T.~Heimburg.
	\newblock 2016.
	\newblock Solitary electromechanical pulses in lobster neurons.
	\newblock Biophys.\ Chem. 216:51--59.
	
	\bibitem{Heimburg2005c}
	Heimburg, T., and A.~D. Jackson.
	\newblock 2005.
	\newblock On soliton propagation in biomembranes and nerves.
	\newblock Proc.\ Natl.\ Acad.\ Sci.\ USA 102:9790--9795.
	
	\bibitem{Lin2000}
	Lin, A.~L., A.~Hagberg, A.~Ardelea, M.~Bertram, H.~Swinney, and E.~Meron.
	\newblock 2000.
	\newblock Four-phase patterns in forced oscillatory systems.
	\newblock Phys.\ Rev.\ E 62:3790--3798.
	
\end{thebibliography}
\small{
	
}

\clearpage
\appendix
\renewcommand{\thefigure}{A\arabic{figure}}
\setcounter{figure}{0}

\section{Calculating K\slash g for the ideal gas}
\label{calculatingkgfortheidealgas}

\renewcommand{\theequation}{A.\arabic{equation}}
\setcounter{equation}{0}

\begin{figure}[htbp]
\centering
\includegraphics[width=169pt,height=73pt]{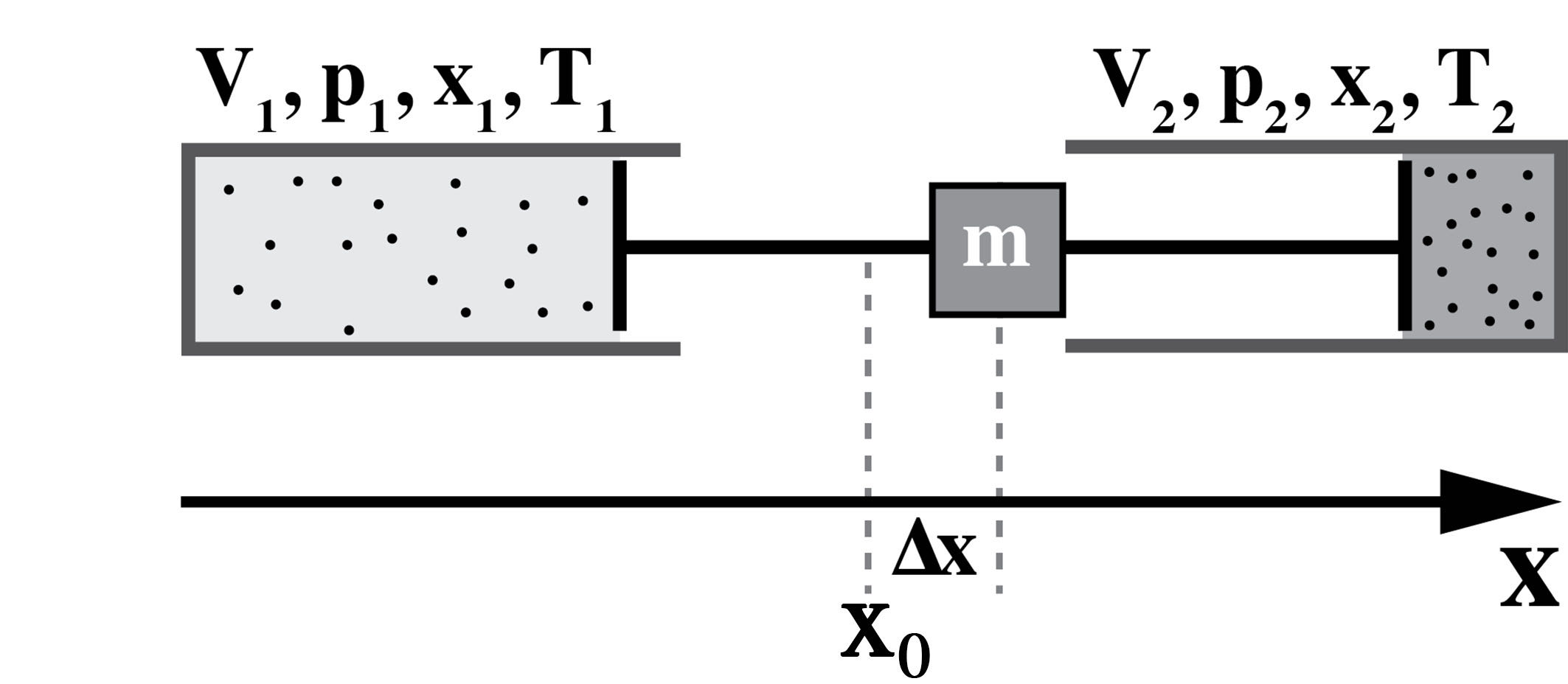}
\caption{Two coupled gas containers (adapted from \cite{Heimburg2017}).}
\label{figure_a1}
\end{figure}

In \cite{Heimburg2017} we showed that in two coupled gas containers (which we consider as a single spring with equilibrium position $x_0$, see Fig. \ref{figure_a1}) containing $N$ particles each, the spring constant of both containers combined upon adiabatic compression is given by
	\begin{equation}\thickmuskip=0mu
\label{eq:A0.1}
K=\frac{10}{3x_0^2}NkT_0
\end{equation}

where $x_0$ is the equilibrium position of the mass, and $T_0$ is the temperature of the two gas containers for $x=x_0$. The change in extension shall be given by $\Delta x$ such that the first container has an extension $x_0+\Delta x$ and the second one has $x_0-\Delta x$.

The change in entropy upon isothermal compression is
	\begin{align}\thickmuskip=0mu
\label{eq:A0.2}
\Delta S&=-\frac{1}{2}g \Delta x^2\qquad\mbox{and}\nonumber\\
\Delta S&=Nk\left[\ln\frac{x_0+\Delta x}{x_0}+\ln\frac{x_0-\Delta x}{x_0}\right]\approx -Nk\left(\frac{\Delta x}{x_0}\right)^2\nonumber\\
&\longrightarrow g=\frac{2Nk}{x_0^2}
\end{align}
This leads to
	\begin{equation}\thickmuskip=0mu
\label{eq:A0.3}
\frac{K}{g}=\frac{5}{3}T_0
\end{equation}

\newpage

\end{document}